\documentclass[fleqn,usenatbib]{mnras}

\usepackage{newtxtext,newtxmath}

\usepackage[T1]{fontenc}

\DeclareRobustCommand{\VAN}[3]{#2}
\let\VANthebibliography\thebibliography
\def\thebibliography{\DeclareRobustCommand{\VAN}[3]{##3}\VANthebibliography}

\usepackage{graphicx}	
\usepackage{amsmath}	
\usepackage{xcolor, wasysym}

\usepackage[normalem]{ulem}

\title[Equatorial ionosphere using the GMRT]{Study of the equatorial ionosphere using the Giant Metrewave Radio Telescope (GMRT) at sub-GHz frequencies}

\author[S. Mangla et al.]{
Sarvesh Mangla,$^{1}$\thanks{E-mail: mangla.sarvesh@gmail.com ; phd1801121006@iiti.ac.in}
Abhirup Datta,$^{1}$
\\
$^{1}$Department of Astronomy, Astrophysics and Space Engineering, Indian Institute of Technology Indore, Madhya Pradesh, 453552, India
}
\date{Accepted XXX. Received YYY; in original form ZZZ}

\pubyear{2022}

\begin{document}
\label{firstpage}
\pagerange{\pageref{firstpage}--\pageref{lastpage}}
\maketitle

\begin{abstract}
Radio interferometers, which are designed to observe astrophysical objects in the universe, can also be used to study the Earth's ionosphere. Radio interferometers like the Giant Metrewave Radio Telescope (GMRT) detect variations in ionospheric total electron content (TEC) on a much wider spatial scale at a relatively higher sensitivity than traditional ionospheric probes like the Global Navigation Satellite System (GNSS). The hybrid configuration of the GMRT (compact core and extended arms) and its geographical location make this interferometer an excellent candidate to explore the sensitive regions between the northern crest of the Equatorial Ionization Anomaly (EIA) and the magnetic equator.
For this work, a bright radio source, 3C68.2, is observed from post-midnight to post-sunrise ($\sim$\,9 hours) to study the ionospheric activities at solar-minima. This study presents data reduction and processing techniques to measure differential TEC ($\delta\rm{TEC}$) between the set of antennas with an accuracy of $1\times10^{-3}$ TECU. Furthermore, using these $\delta\rm{TEC}$ measurements, we have demonstrated techniques to compute the TEC gradient over the full array and micro-scale variation in two-dimensional TEC gradient surface. These variations are well equipped to probe ionospheric plasma, especially during the night-time. Our study, for the first time, reports the capability of the GMRT to detect ionospheric activities. Our result validates, compared to previous studies with VLA, LOFAR and MWA, the ionosphere over the GMRT is more active, which is expected due to its location near the magnetic equator.
\end{abstract}




\section{Introduction}
The Earth's ionosphere limits the low-frequency ($<$\,1\,GHz) observation for ground-based radio-telescopes. The three-dimensional ionosphere consists of partially ionized plasma, which introduces systematic effects \citep[see][]{Thom2001isra.book.....T, Mangum2015PASP..127...74M} such as refraction, reflection, propagation delay, and Faraday rotation.
Interferometers such as GMRT\footnote{Giant Meterwave Radio Telescope, Pune India \url{http://www.gmrt.ncra.tifr.res.in}}, VLA\footnote{Very Large Array, New Mexico \url{https://science.nrao.edu/facilities/vla}}, LOFAR\footnote{LOw Frequency ARray, Netherlands \url{https://www.astron.nl/telescopes/lofar/}}, MWA\footnote{The Murchison Widefield Array \url{https://www.mwatelescope.org/}} and future instruments such as the Square Kilometre Array \citep[SKA;][]{Dewdney2009IEEEP..97.1482D} are all affected by the ionosphere similarly. As the new generation low-frequency radio telescopes are in operation, they can probe the ionosphere in unprecedented detail. The calibration of these ionospheric effects especially at the low-frequency radio observation, becomes a challenging task. \par
\begin{figure*}
    \centering
    \includegraphics[scale=0.38]{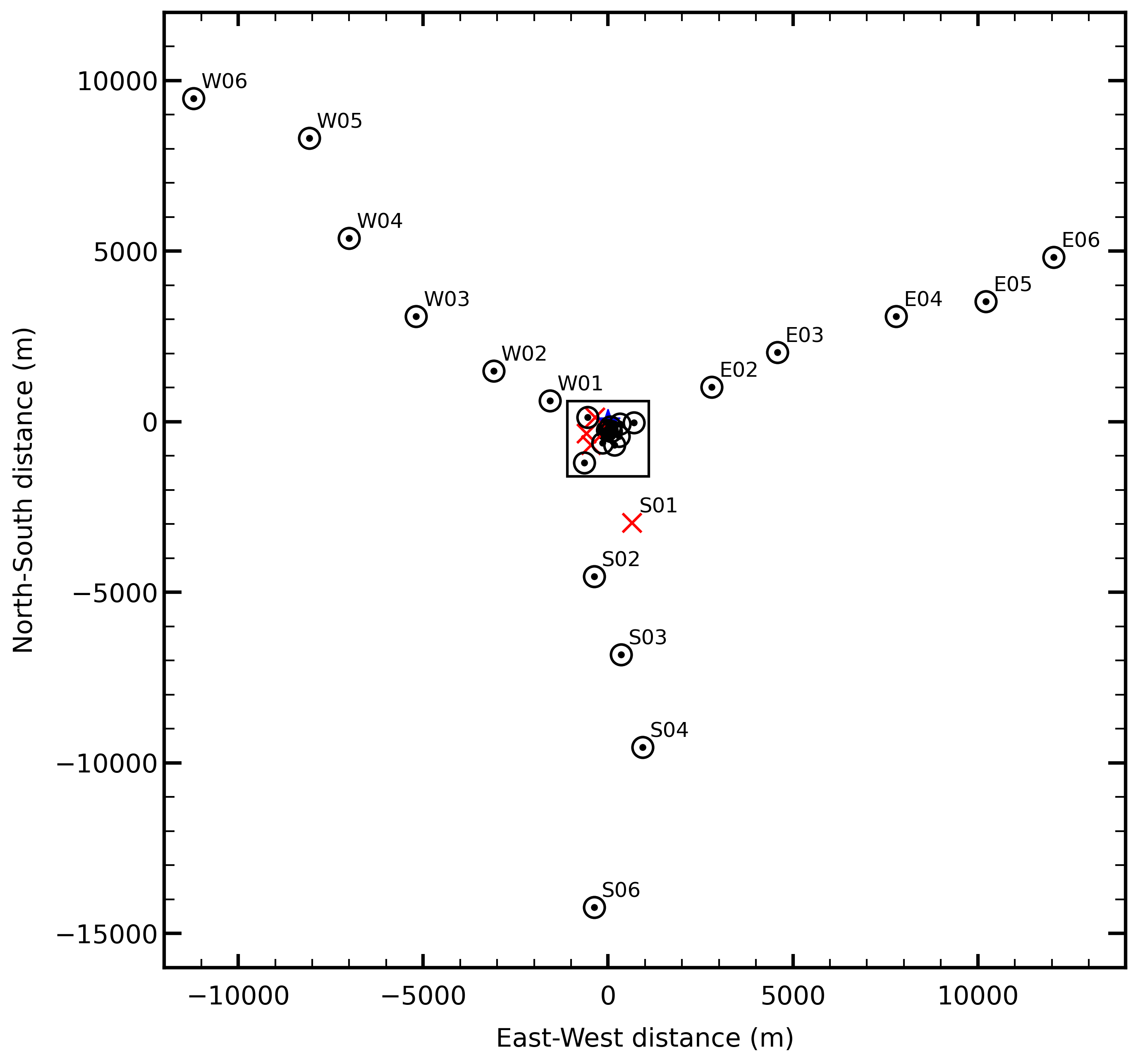}
    \includegraphics[scale=0.38]{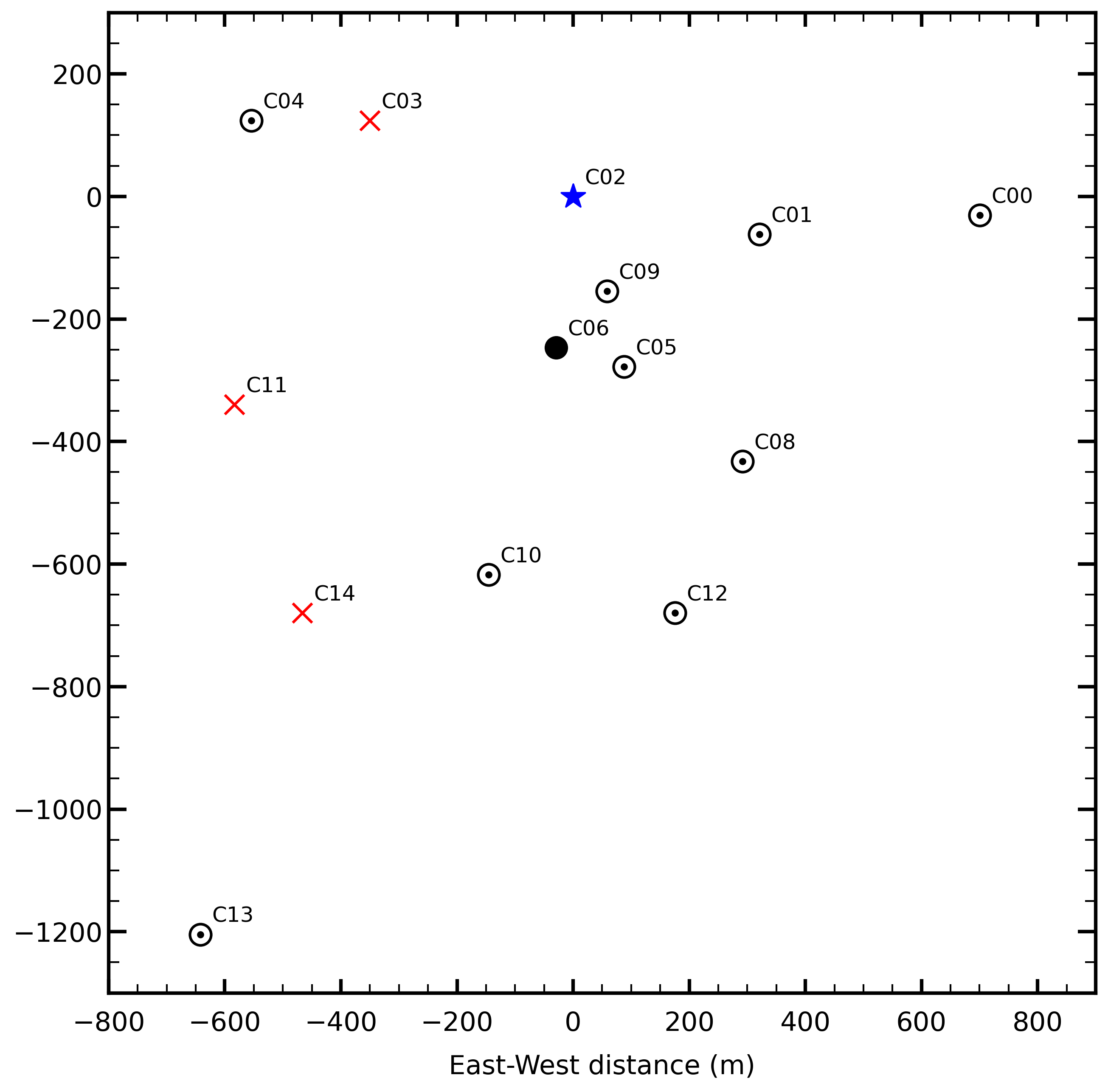}
    \caption{Left Panel: shows the layout of the full GMRT array during the observation run of the astronomical source (3C68.2), Right Panel: shows the zoomed in version of the central box from the left panel. Blue star (``C02'') is the array centre; the reference antenna (``C06'') is highlighted in the black circle. The red crosses are bad antennas during the observation.}
    \label{fig:gmrt_array}
\end{figure*}
The Earth's ionosphere is directly dependent on solar and geomagnetic activity and continues to vary due to tides and atmospheric gravity waves. It also depends on the latitudinal variation; the electron density keeps decreasing as one ascends towards the latitudes of both hemispheres, showing a maximum around the geomagnetic equator \citet{Opio2015AdSpR..55.1640O}. The low-latitude extending up to $\pm20^\circ$ magnetic latitudes consists of the Equatorial Ionization Anomaly (EIA) region, where the ionosphere varies at dawn and has unanticipated changes during the night-time. The main drivers of hazardous ionospheric variation in the EIA region are equatorial plasma bubbles (EPB), which are considered to be initiated by gravity waves and generate small-scale structures via the Rayleigh-Taylor instability \citep[see][]{Yokoyama2017PEPS....4...37Y}. These activities in the EIA region may disrupt communication and navigation, such as the Global Positioning System (GPS). \par
Interferometers like the VLA ($\sim$\,$34^{\circ}$N) have already been used to probe finer-scale (0.1 mTECU) fluctuation in the ionosphere using a single bright source in the field of view (FoV) \citep[see][and references within]{ Helm2012RaSc...47.0K02H, Jac1992A&A...257..401J}. It explores the phenomenology of the ionosphere by performing Fourier analysis of measured phases for individual antennas \citep[see][]{Jac1992P&SS...40..447J, Helm2012RaSc...47.0L02H}. Widefield interferometers that can observe many celestial radio sources simultaneously, uses positional offset to study the ionosphere \citep[see][]{Cohen_2009}. \citet{Helmboldt2012RaSc...47.5008H} detected groups of wavelike disturbances by conducting a spectral analysis of 29 bright source position shift in a single FoV using VLA. \par
\citet{mev16} reported that ionospheric irregularities ($\delta\rm{TEC}$) were anisotropic during the night-time with an accuracy of $1\times10^{-3}$ TEC using LOFAR (situated in mid-latitude). Using data only from night-time observations during the winters of 2012-2013, a power-law behaviour was also observed over the long-range baseline lengths, explaining the Kolmogorov turbulence in the ionosphere. Later, \citet{gasp18} showed that at ultra-low frequencies, the visibility amplitudes are corrupted by scintillation, due to which an average of 30 percent of the data is lost during the night-time (compared to the day-time). Observations are therefore recommended during day-time hours, especially for the LOFAR-EoR project. Recently, \citet{Rich2020JSWSC..10...10F} with a combined analysis of Global Navigation Satellite System (GNSS), ionosondes and LOFAR reveals large-scale travelling ionospheric disturbances (TIDs) and small-scale TIDs travelling perpendicular to each other. These cause instabilities which breakdown large-scale structures into smaller scales during a weak sub-storm at high latitudes. \par
\citet{Loi2015RaSc...50..574L} studied the ionosphere by power spectrum analysis of ionospheric fluctuations (computed using positional offsets of radio sources) using MWA ($\sim$\,$26^{\circ}$S) data. This analysis reveals a range of wavelike phenomena in which some fluctuations show characteristics of TIDs. Recently, \citet{Helm2020RaSc...5507106H} constructed images of ionospheric structures using 200\,hrs of GLEAM survey data by MWA \citep[see][]{Wayth2015PASA...32...25W}. Spectral analysis of these images revealed distinct features in the night-time ionospheric activity. \par
Although the Earth's ionosphere is studied using these telescopes, the studies are limited to their local ionospheric region due to geographical location constraints. This makes the GMRT a unique telescope to study the ionosphere, due to its favorable geographic location (latitude = $19^{\circ} \, 05' \, 35.2''$ N and longitude = $74^{\circ} \, 03' \, 01.7''$ E) which is the geophysically sensitive region between the magnetic equator and the northern crest of the EIA \citep[see][]{Appl1946Natur.157..691A} in the Indian longitude sector. Like VLA, GMRT is also well equipped to study the ionosphere because antennas have a similar ``Y'' shaped configuration, allowing us to explore the ionospheric variations in three different directions. The configuration of GMRT consists of 14 antennas in the central square of $\rm 1.4\times1.4\,km^2$, and 16 antennas along three arms, each of approximately 14\,km, which gives us the ability to explore the ionospheric variations over a broader range of scales mentioned in \citet{Lonsdale2005ASPC..345..399L}. 
This makes the GMRT an astute telescope to study the EIA region due to its geographical location and configuration. This is the first work with the GMRT to study the Earth's ionosphere. \par
This paper is structured as follows, in Sect. (\ref{sec:method}) quantitatively describes the ionosphere's effect on electromagnetic radiation at low radio frequencies and how this affects the data obtained from any radio telescope. In Sect. (\ref{sec:data}) the data selection with GMRT and post-processing has been done to get phase information, which will be used to study the ionosphere. Following this in Sect. (\ref{sec:tec_gradient}), we describe the process to get the ionospheric phase and mitigate the instrumental noise and convert the resulting differential phase to differential TEC. In Sect. (\ref{sec:tec_grad_general}), we describe polynomial based and arm-based methods to compute the TEC gradient over the full array and along each arm direction, respectively. Finally, we conclude our results and discuss the prospects for future work in Section (\ref{sec:results}).
\section{Radio Interferometric observation through the Ionosphere}
\label{sec:method}
A radio interferometer is an array of elements (antennas) that measure the ``spatial coherence function'' at the pair of antennas pointed towards an astronomical object \citep[see][]{Cornwell1999ASPC..180..187C}. The observed visibility for a pair of antenna or ``baseline'' is given by:
\begin{eqnarray}
    \label{eq:visibility}
    V_{\rm \nu}(u,v,w) = \int \int I_{\rm \nu}(l,m,n) e^{-2 \uppi i (ul+vm+nw)} d \Omega
\end{eqnarray}
where $\Omega$ denotes the solid angle, \textit{$I_{\rm \nu}$} is the intensity observed on the sky at frequency ($\nu$) at a position given by the direction cosines $l$, $m$, and $n = \sqrt{1 - l^2 - m^2}$, which are measured along the line of sight to the source. The spatial frequencies are defined as $u$, $v$ and $w$. The difference of the position vector between the two antennas is normalised by the observing wavelength and their directions, which is based on the position of the object being observed in the sky. This coordinate system is defined so that $w$ is along the line of sight to the source in the sky, $u$ and $v$ correspond to the east-west and north-south direction, respectively. These frequencies change for a pair of antennas as the object changes its position in the sky.
Interferometers work as ``fringe-stopped'', which means that the observed visibility is multiplied by an additional factor of $e^{2 \uppi i w}$, such that at the phase centre the phase is zero \citep[see][]{thompson_book}. The above equation (\ref{eq:visibility}) becomes a function of $u$ and $v$, which can easily be converted to measure the intensity map using standard numerical methods.
The ionosphere limits our ability to explore the sky at sub-GHz frequencies by introducing systematic effects in which propagation delay is the most dominant term for interferometric observation. This delay introduces an extra phase term, which is caused due to varying refractive index ``n'' of the ionospheric plasma along the line of sight \citep[see][]{intema_thesis}. Using Snell's Law, total propagation delay at frequency $\nu$, integrated along the line of sight results in a phase rotation, which is given by:
\begin{eqnarray}
    \label{eq:phi_total}
    \phi_{\rm ion} = - \frac{2 \uppi \nu}{c} \int (n - 1)dx
\end{eqnarray}
where c is the speed of light in vacuum. A constant phase gradient would cause the spatial shift of the sources in the image plane compared to the actual positions in the sky. The problem becomes more severe as ``n'' strongly depends on the position and time of the observation. For a cold, collision-less plasma without a magnetic field with frequencies $\nu \gg \nu_p$ (plasma frequency = 1-10MHz for ionosphere) equation (\ref{eq:phi_total}) can be expanded (see e.g. \citealt{Thom2001isra.book.....T}) using Taylor series approximation. The first term (the dominant term) is associated with a dispersive delay, proportional to the TEC along the line of sight, which is given by: 
\begin{eqnarray}
    \label{eq:phi}
    \phi_{\rm ion} = \frac{e^2}{4 \uppi \epsilon_0 c m_e \nu}\int n_e(s)ds 
\end{eqnarray}
\noindent This can be approximated as 
\begin{eqnarray}
    \label{eq:phi2tec}
    \phi_{\rm ion} = 84.36 \left ( \frac{\nu}{\mbox{\scriptsize 100 MHz}} \right )^{-1} \left ( \frac{\mbox{\scriptsize TEC}}{\mbox{\scriptsize 1 TECU}} \right ) \mbox{ radians}
\end{eqnarray}
where the integral over $n_{e}$ (free electron density) is referred to as TEC along the line of sight, s denotes the path-length through the ionosphere along the line of sight, $\epsilon_0$ is the electric permittivity of a vacuum, c is the speed of light, and $m_e$ is the mass of an electron. The TEC unit (TECU) is $10^{16}$ electrons $\rm m^{-2}$, which is the typically observed value at zenith during night-time. \par
One can easily infer from equations (\ref{eq:phi} and \ref{eq:phi2tec}) that the computed phase correction terms can be stated as the difference between the TEC along the line of sight of a given antenna to that of the selected reference antenna. As the first-order terms are not being appropriately calibrated before the Fourier inversion process to make an image of the sky, the apparent position of the sources in the image plane may be affected. If higher-order phase terms begin to dominate, the sources may appear distorted or even disappear in the image plane. The higher-order terms can be ignored for frequencies higher than a few hundred megahertz \citep[see][]{thompson_book}. Therefore, calibrating the extra phase due to the ionosphere is necessary to locate astrophysical sources accurately. This phase term can give an understanding of the dynamics of the ionosphere. \par
Typically, the initial amplitude and phase calibration for instrumental effects is followed up with few rounds of imaging and ``self-calibration'' \citep{book1989ASPC....6...83F}. Self-calibration helps in removing any extra phase errors to improve the quality of the resulting image. This procedure involves solving for complex gains solutions as a function of time, chi-squared minimisation based on the assumed source model obtained through an initial image of the source. These complex gain solutions are computed between two antenna elements or a baseline. Choosing one reference antenna from the array (preferably from the centre) and setting the phases of its complex gain to zero, one can calculate the computed gain correction relative to this reference antenna. For N antennas or N(N-1)/2 baselines, the gain solutions can be computed relatively for a short period, depending upon the source's brightness. Generally, this method is used to produce an image, deconvolve it again to make a better sky model and repeat the procedure until it converges with the sky model \citep[see][and references therein]{Cornwell1999ASPC..180..187C}. \par
Considering the size of an array (baseline as small as $\sim$\,100\,m to as large as $\sim$\,25\,km) and a short period for ``self-calibration'' (usually $\sim$\,1\,minute or around tens of seconds for bright sources), these interferometers are capable of studying small TEC variations (such as $\delta\rm{TEC}$ as small as 0.1\,mTECU \citep[see][]{Helm2012RaSc...47.0K02H}, when observing a very bright source). These instruments are better to study small-scale fluctuations than many other instruments available like GNSS because of their sensitivity. \par
\section{Observations and Data Analysis}
\label{sec:data}
For this study, we have observed a bright radio source, 3C68.2 (total intensity $I_\nu \sim 3$~Jy at the observing frequency of 610\,MHz) located towards the north of the GMRT array, i.e. towards the northern crest of EIA. As per the observation strategy, the FoV containing a bright compact source at the centre, which is at least ten times brighter than the next brighter source in FoV makes it suitable for direction-independent self-calibration procedure.
\begin{table}
\centering
    \caption{The observation summary}
    \label{tab:obs_sum}
    
    \begin{tabular}{|c|c|}
    \hline
        Project code & 22$\_$064 \\
        Observation date & 06 Aug 2012 \\
        Start and End Time of Observation (IST) & 01:35\,AM to 11:10\,AM \\
        Bandwidth & 16.67 MHz \\
        Central Frequency & 610 MHz \\
        Channels & 128 \\
        Frequency resolution & 130 kHz \\
        Integration time & 0.5 s \\
        Correlation & RR \\
        Working antennas & 26 \\
        Calibrator (Flux $\&$ Phase) & 3C48 \\
        Target source & 3C68.2 \\
        RA (J2000) & 02:34:23.81 \\
        DEC (J2000) & +31.34317.03 \\
        Total on-source time & $\sim$ 9 h \\
    \hline
    \end{tabular}
\end{table}
\subsection{Observing with the GMRT}
\label{sec:gmrt_observation}
The GMRT \citep{swarup_GMRT} is one of the largest and fully operational sensitive telescopes to low radio frequencies. The GMRT consists of 30 dishes, each 45\,m diameter spanning over 25\,km providing a total collecting area of about 30,000\,$\rm m^2$ at metre wavelengths, with a moderately good angular resolution ($\sim$\,arcsec). Out of 30 antennas, 14 antennas are randomly distributed in a central square, which is 1.4\,$\times$\,$1.4\,\rm km^2$ in extent. The other 16 antennas are placed along three arms, each about 14\,km long in an approximately `Y' shaped configuration. The array layout is shown in Fig~\ref{fig:gmrt_array}. We carried out an observation of 3C68.2 with GMRT at 610 MHz approximately for 9 hours on 5-6 August 2012 (Proposal code: 22\_064). The observation summary is presented in Table \ref{tab:obs_sum}. At the time of the observation, there was a moderate amount of geomagnetic activity ($K_p$ index $\sim$ 1-3) \footnote{OMNIWeb \url{https://omniweb.gsfc.nasa.gov/ow.html}\label{footnote omniweb}} and low solar activity (F10.7 = 137.9 SFU; 1SFU = $\rm 10^{-22}\,W\,m^{2}\,Hz^{-1}$)\textsuperscript{\ref{footnote omniweb}}. During the observation, out of the 30 antennas, C11, C14, S01 were not operational. The dataset contains each scan of nearly one hour, except scan 4, which is nearly four hours long during post-sunrise hours. For this dataset, the total bandwidth was 16.67 MHz of RR polarisation. It comprises of 128 channels of channel-width 130.2 kHz and time-resolution of 0.5 seconds. \par
\begin{figure*}
    \centering
    \includegraphics[scale=0.50,angle=0.0]{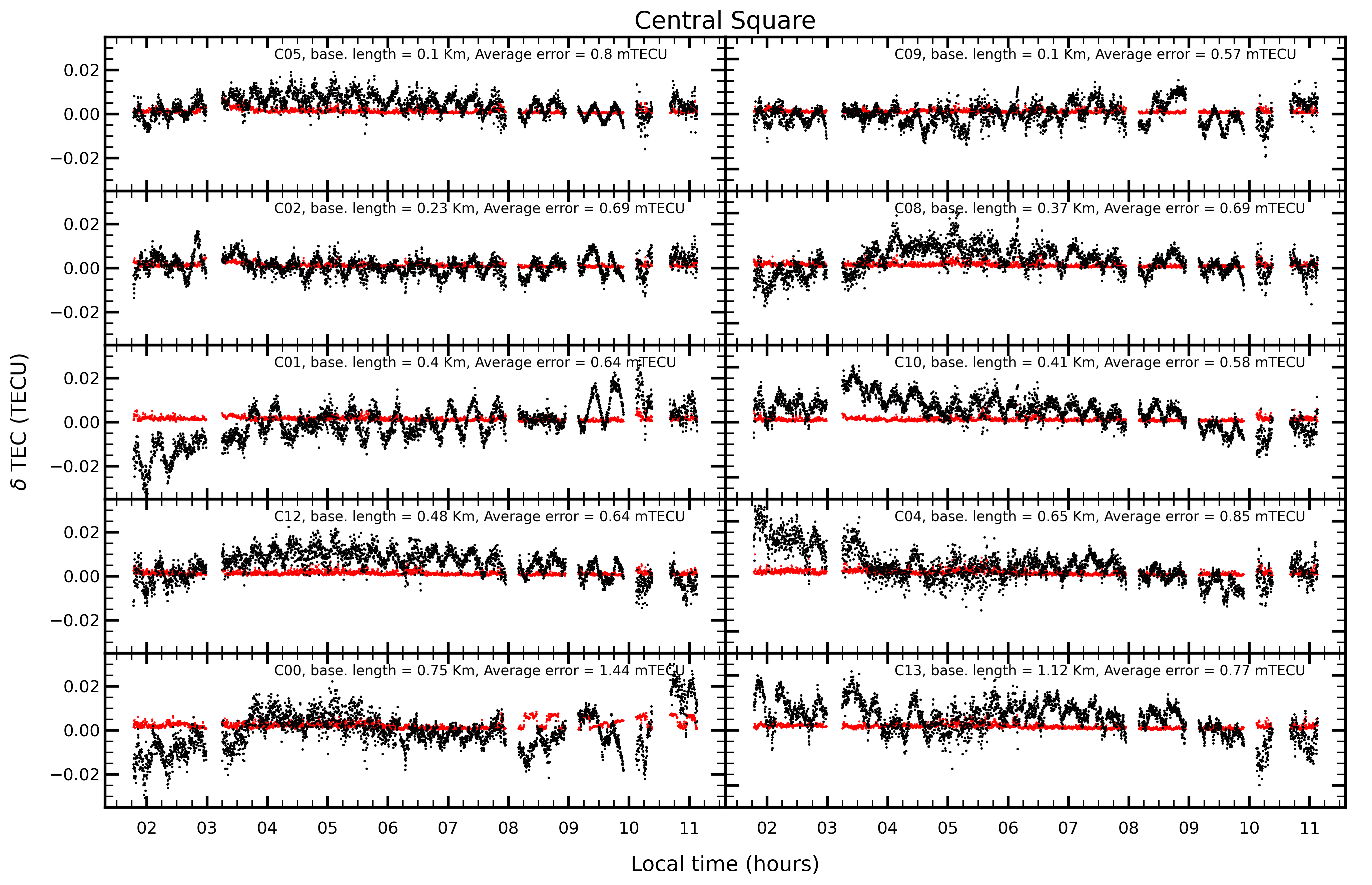}
    \caption{Differential TEC ($\rm \delta TEC$) for each antenna in the compact Central Square of the GMRT is plotted as a function of time. This $\rm \delta TEC$ was estimated along antenna's line of sight with respect to the reference antenna line of sight. In each panel, the estimated uncertainty is also plotted in red.}
    \label{fig:central_sq}
\end{figure*}
One must deal with a significant amount of radio-frequency interference (RFI) during the processing of low-frequency interferometric data from the GMRT. Ionospheric conditions may differ from one part to another part of the FoV, which require special calibration techniques. Imaging the entire FoV regardless of the source(s), it is crucial to minimise noise spikes or residual sidelobe in the image plane.
\subsection{RFI mitigation}
\label{sec:rfi_mitigation}
RFI limits the sensitivity of radio observations by increasing the system noise and corrupting the calibration solutions. It also restricts the available frequency bandwidth. The effect is particularly strong at frequencies below 600 MHz at GMRT. We have used an auto-flag algorithm, Common Astronomy Software Applications\footnote{Common Astronomy Software Applications (\url{https://casa.nrao.edu/})} task \texttt{TFCROP}, which detects outliers on the 2D time-frequency plane for each baseline. The algorithm iterates using a sliding window in time defined by the user to construct a clean band-shape (without RFI) template across the base of RFI spikes. In each iteration, it computes the standard-deviation between the data and the fit, and flags the values beyond N-sigma for time and M-sigma for frequency domain separately. We have used 4-sigma for the time domain, and 3-sigma in the frequency domain optimised for strong narrow-band RFI. We have also used mode \texttt{EXTEND} in \texttt{CASA} to flag the data if \texttt{TFCROP} flags 50$\%$ data in the time and frequency bins. \par
We have also applied the auto-flag algorithm, \texttt{RFLAG}, on the calibrated data, where the data is iterated through in segments of time, local RMS and median RMS of the real and imaginary parts of the visibilities across channels and across a sliding time window. The deviation of the local RMS from this median value has been calculated. If local RMS is larger than ten times the median value of deviation, the data is flagged. The bulk of flagging is done using \texttt{RFLAG} on all data uniformly for direction-independent calibration. Rest is minor iterative flagging during calibration steps.
\subsection{Calibration - Flux, Phase and Bandpass}
\label{sec:direction_inde_cal}
After flagging of the spurious signal present in the dataset, we do a direction-independent calibration using standard tasks within the \texttt{CASA} package. \par
We use 3C48 as flux density and bandpass calibrator as well as phase calibrator. The \citet{SetJy} model is used to set the flux value of 3C48 using \texttt{SETJY} task in \texttt{CASA}. The first step in the calibration is to do a bandpass calibration using the \texttt{CASA} task \texttt{BANDPASS} to determine the bandpass response as a function of frequency for each antenna. This task is performed to correct the frequency-dependent instrumental response. One chooses the entire band for this task, but the first and last few channels can be excluded where the response drops substantially. Following this, we calculate the gain and phase variation as a function of time on a 10\,s time-scale, using the \texttt{CASA} task \texttt{GAINCAL} for calibrator 3C48. This task determines the time-dependent antenna-based variation needed to fit the observed visibilities to the model and flags any spurious time-intervals for a given antenna. \par
As described in Section (\ref{sec:method}), an interferometer is only sensitive to measure relative phases; thus, one chooses a reference antenna whose phase is subsequently set to zero. Generally, the RFI effects tend to be the worst for the central antenna in the array; thus, we use C06 as our reference antenna (see Fig.\ \ref{fig:gmrt_array}). \par
\begin{figure}
    \includegraphics[scale=0.45,angle=0.0]{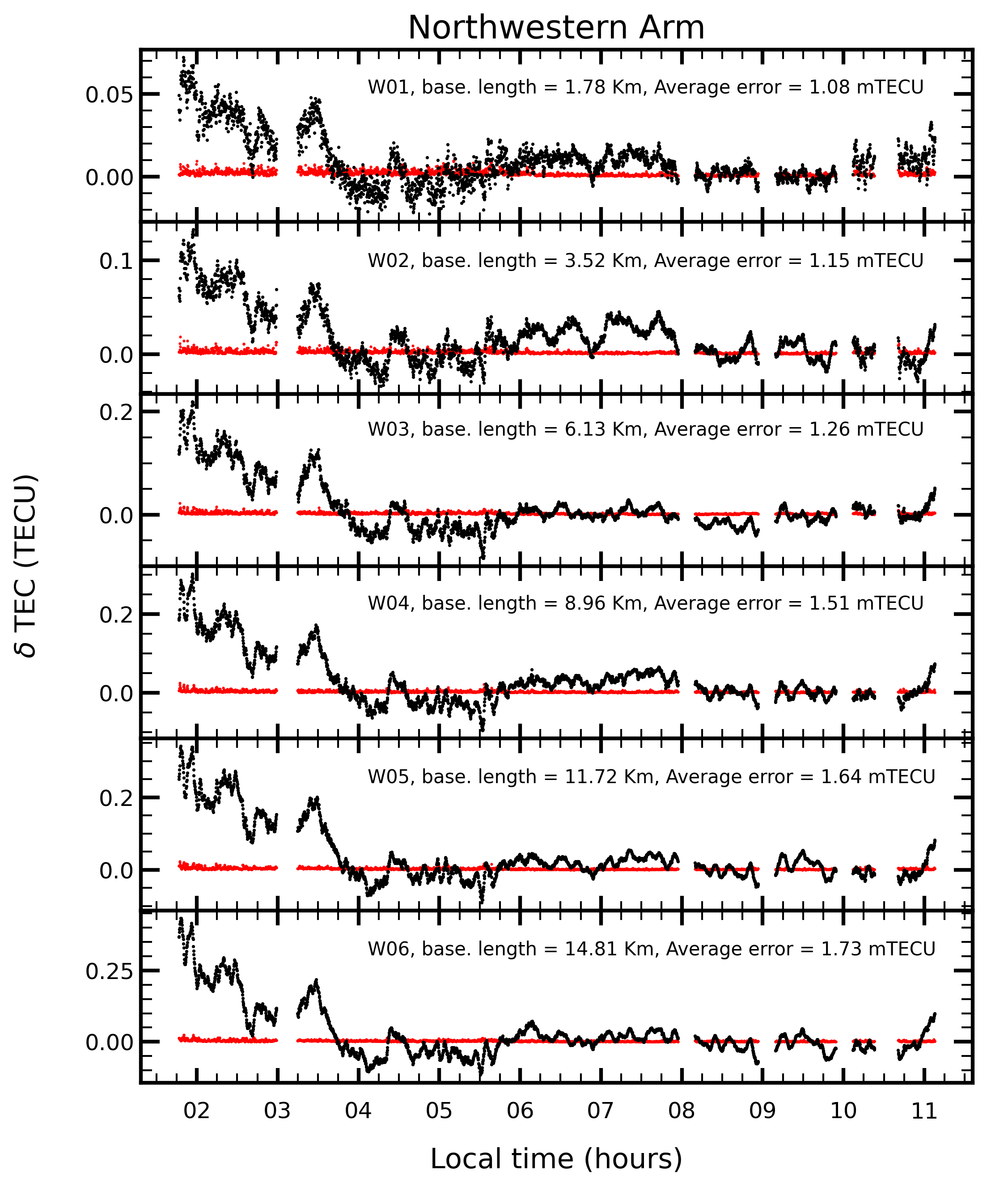}
    \caption{Same as Fig.\ \ref{fig:central_sq} but along the North-Western arm of the GMRT.}
    \label{fig:northwestern}
\end{figure}
We use the calibrator source 3C48 as flux density, phase gain and band-pass calibrations, as it is near the target source (3C68.2). During the calibration, bad data is flagged for an antenna with significant error in complex gain solutions. Some baselines are also flagged for specific scans in the data. After applying calibration and RFI mitigation to the target field, $32.6\%$ of data is flagged. \par
Dual-frequency observation is beneficial to obtain the ionospheric phase solutions by removing the instrumental phase error as the ionospheric phase scales with wavelength (see equations \ref{eq:phi} and \ref{eq:phi2tec}). However, the instrumental phase does not necessarily behave in the same way; therefore, in order to remove the instrumental phase error, we split the data centered around two frequencies, 605 MHz and 614MHz. \par
\begin{figure}
    \includegraphics[scale=0.45,angle=0.0]{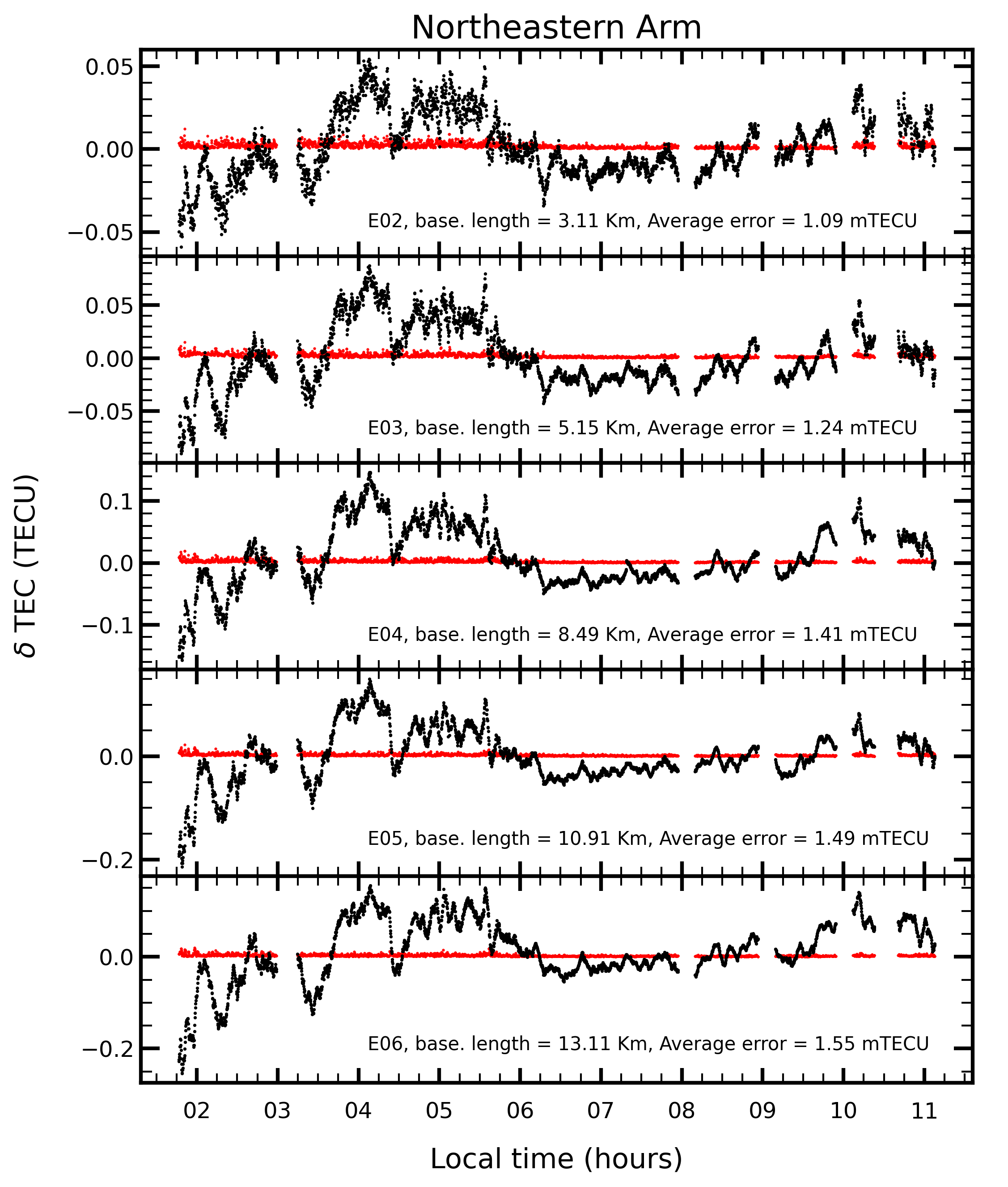}
    \caption{Same as Fig.\ \ref{fig:central_sq} but along the North-Eastern arm of the GMRT.}
    \label{fig:northeastern}
\end{figure}
\subsection{Imaging and Self-calibration} 
At $610$~MHz, the GMRT primary beam has a large FoV of ($\rm 0.75^{\circ} \times 0.75^{\circ}$). We implement wide-field imaging tools within \texttt{CASA} by using gridmode=`widefield' in the \texttt{CASA} task \texttt{CLEAN}. This is crucial to perform widefield imaging using ``wprojection'' in order to take the effects of non-coplanar array geometry and wide FoV of the GMRT at these frequencies. We use Briggs robust parameter $\rm -1$ as these shifts slightly towards uniform weighting. \par
At any low-frequency radio observation, the sky dominates the system temperature. In order to increase the signal-to-noise ratio, we average the data over several timestamps while finding complex gain solutions. However, the ionosphere tends to vary quickly with time, which prevents us to average the data over larger time intervals. To ensure accurate calibration for all antennas and ionosphere conditions, the calibration is performed at a shorter (10\,sec) time interval (raw data is taken at 0.5\,sec cadence), using the 3C68.2 model visibilities. The self-calibration (phase only) on the target is performed using the \texttt{CASA} task \texttt{GAINCAL}. This task flags spurious solutions for a subsequent time interval and antennas based on two criteria: the minimum signal-to-noise ratio (we take two because of the short time interval) and the minimum number of antennas (we take six). Bad solutions are obtained for antenna ``C03'' and flagged manually. \par
\section{Ionospheric Information from the Radio Observations}
\label{sec:tec_gradient}
\subsection{Additional phase error due to the ionosphere}
\label{sec:processing_phase_correction}
To estimate the additional phase introduced due to the radio signal passing through the Earth's ionosphere, we follow the procedure outlined in \citet{Helm2012RaSc...47.0K02H}. After calculating the phase terms from \texttt{CASA}, several steps are performed on phase data to get the ionospheric information. The phase terms contain other effects along with the ionospheric contribution, which is highest in low-frequency regimes. \par
The phase difference between the two antenna elements \citep[see][]{Intema2009A&A...501.1185I,Helm2012RaSc...47.0K02H} is given by 
\begin{eqnarray}
        \Delta\phi = \Delta\phi_{\rm ion} + \Delta\phi_{\rm instr} + \Delta\phi_{\rm amb} + \Delta\phi_{\rm sour} 
    \label{eq:Total_phase_diff}
\end{eqnarray}
where $\Delta\phi_{\rm ion}$ represents the difference in the ionospheric phases of the two antennas along the line of sight given by equation (\ref{eq:phi2tec}), $\Delta\phi_{\rm instr}$ denotes the difference in the instrumental effects between the two antennas, $\Delta\phi_{\rm amb}$ is the contribution from $2\uppi$ ambiguities and $\Delta\phi_{\rm sour}$ is the phase difference from the observed source structure. \par 
One may easily remove $\Delta\phi_{\rm amb}$ by having a short time sampling to ``unwrap'' the phases, and $\Delta\phi_{\rm sour}$ by dividing the observed visibilities by a model of ``3C68.2''. The observations at 605 MHz and 614 MHz is used to remove the instrumental effects as the ionospheric phase scales with wavelength. The instrumental phase includes the error associated with the delays, the actual source position and the offset between antenna line of sight. $\Delta\phi_{\rm instr}$ and $\Delta\phi_{\rm amb}$ are removed by performing further three separate steps. \par
\begin{figure}
    \includegraphics[scale=0.45,angle=0.0]{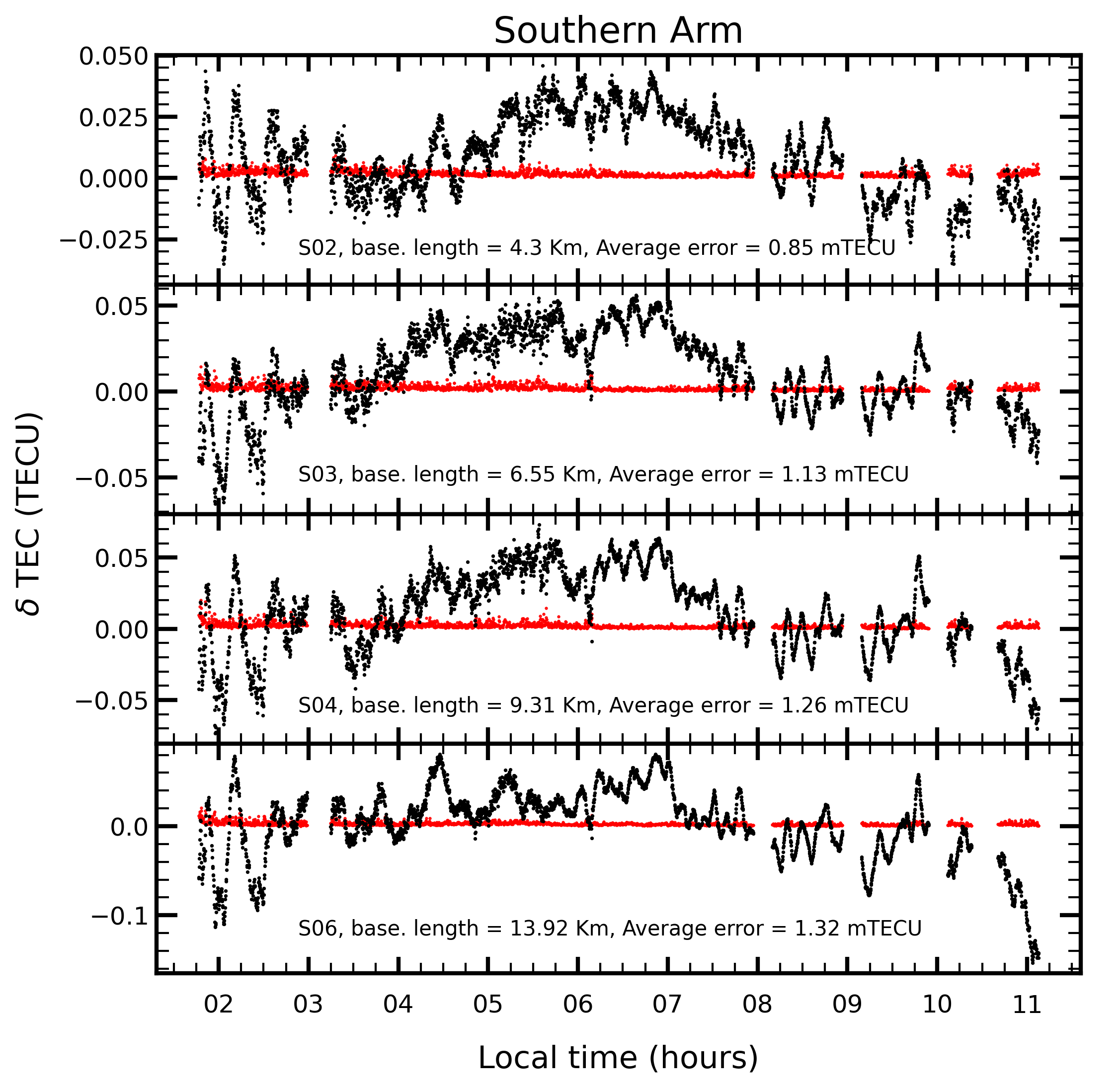}
    \caption{Same as Fig.\ \ref{fig:central_sq} but along the Southern arm of the GMRT.}
    \label{fig:southern}
\end{figure}
\begin{enumerate}
    \item Firstly, the \texttt{CASA} task \texttt{GAINCAL} flags a few time steps that are spurious in nature in some of the antennas. We fill these missing time steps by interpolating (linear interpolation) the real and imaginary components of the complex antenna gains, $g_{\rm A}$, computed by \texttt{GAINCAL} for antennas onto a conventional time-step grid. The phase corrections are recomputed using:
\begin{equation}
    \Delta \phi = \mbox{tan}^{-1}[\mbox{Im}(g_{\rm A})/\mbox{Re}(g_{\rm A})]
\end{equation}
The phase is computed for every time step (2949 steps), which are equally spaced by 10 seconds.
\item Following this, the 10 second time sampling is sufficiently short that the phase, $\phi$, could be ``unwrapped'' in the traditional way as a function of time for each antenna. There are few time steps for some antennas, where the phase correction appears as sharp spikes in the unwrapping process. Few sharp spikes in the unwrapped phase of an antenna is not an issue. However, during the unwrapping process, these spikes may cause a large false phase jump that needs to be taken care of. To improve the phase correction from sudden spikes for a given antenna, we apply an algorithm that calculates the difference between $\mbox{cos}(\phi)$ at a particular time step to the value of the following time step, where $\phi$ is the wrapped phase. Following this, we flag any time step, whose absolute value of the difference is greater than ten times the standard deviation, computed using all-time steps for a particular antenna. These missing time steps are then further filled by the same linear interpolation method using the unwrapped phase data of the un-flagged time steps. As these spikes only occur in 4-5 time steps, linear interpolation is a reasonable method to improve these phase solutions for a given antenna.
\item Further, to mitigate the effects due to the instrument (including errors due to offsets between source position and phase centre), we follow the process defined in \citet{Helm2012RaSc...47.0K02H} as a continuum subtraction process. In this process, one can treat ionospheric fluctuations as superimposed features onto a smooth continuum of instrumental phases, which does not fluctuate as much as ionospheric phases. We have performed our continuum subtraction process for antenna by calculating the difference between the 605\,MHz phase and the 614\,MHz phase scaled by (614/605), so that the ionospheric phases cancel out (as ionospheric phase is $\propto \nu^{-1}$) and provide us with the scaled difference between the 614 and 605 MHz instrumental phases. \par
Because of the positional offset, $\Delta l$ and $\Delta m$ of direction cosines $l$ and $m$, an additional phase of $-2 \uppi (u \Delta l + v \Delta m)$ is also present (see equation \ref{eq:visibility}). As $u$ and $v$ are normalised by the observing wavelength, the phases will be greater for 614 MHz by a factor of 1.015. The offset can arise as different model images are used for each band, and they may not be perfectly aligned. As ionospheric phase is $\propto \nu^{-1}$, we compute $\phi_{605}$ - $\phi_{614}\times (614/605)$ for each antenna as a function of time. This difference is fitted using a linear combination of un-normalized $u$ and $v$ coordinates separately for an one-hour-wide fixed boxcar or the duration of the scan, whichever is shorter. The output between the continuum of instrumental phase and smoothed fitting using $u$ and $v$ coordinates is instrumental noise. The computed instrumental noise is subtracted from respective antenna and band. The remaining part is left as the ionospheric phase term.
\end{enumerate}
\begin{figure}
    \centering
    \includegraphics[scale=0.38]{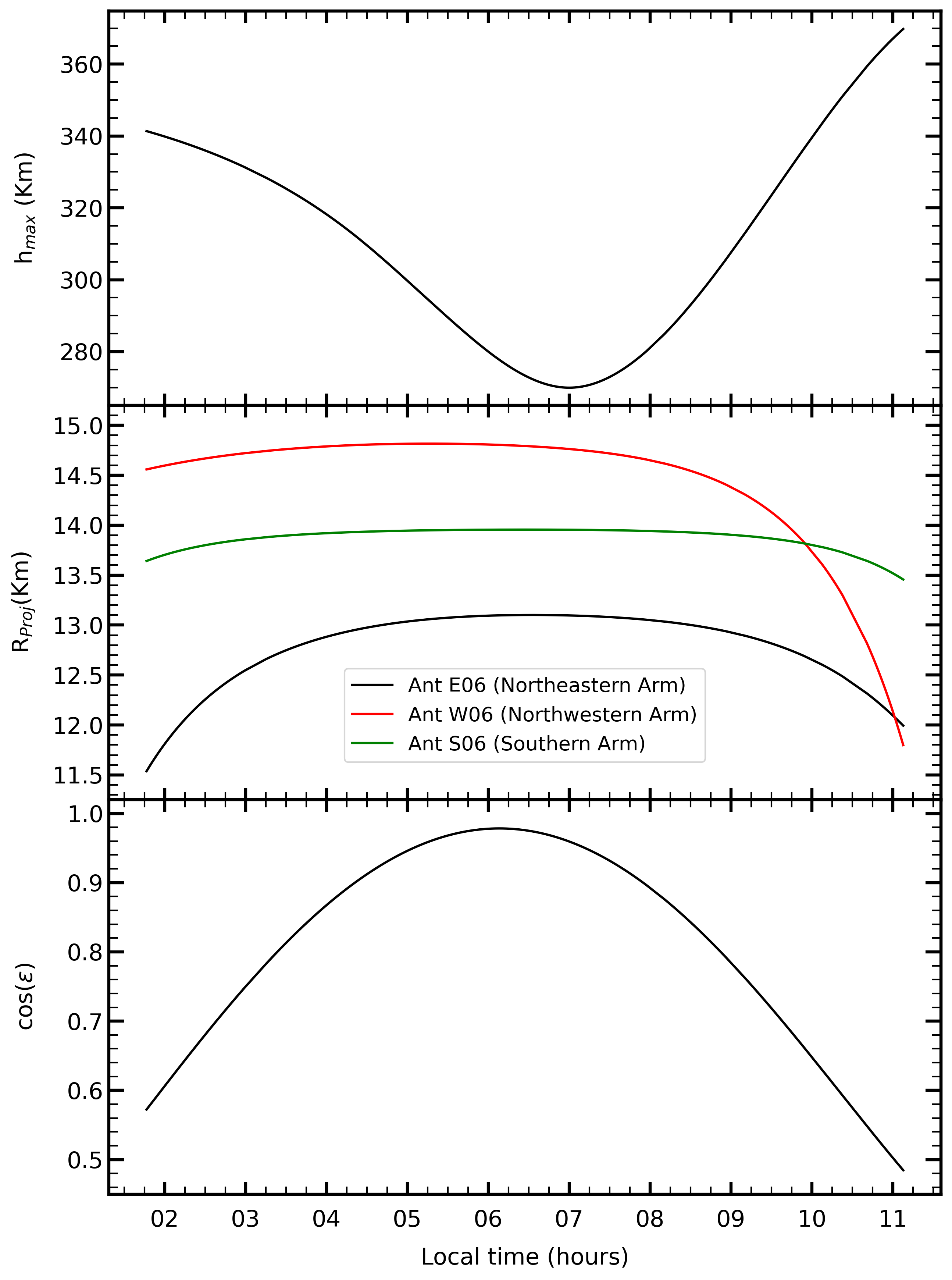}
    \caption{(top) The altitude of maximum electron density is plotted as a function of time computed using the IRI-Plas software at the geographical locations of ionospheric pierce points. (middle) Projected distance for antennas furthest in the arms from the array centre is plotted assuming a thin-shell model at an altitude plotted in figure (top). (bottom) Slant-$\delta\rm{TEC}$ to vertical-$\delta\rm{TEC}$ multiplicative factor is plotted assuming a thin-shell model at an altitude plotted in figure (top).}
    \label{fig:zion}
\end{figure}
\begin{figure*}
    \centering
    \includegraphics[scale=0.5]{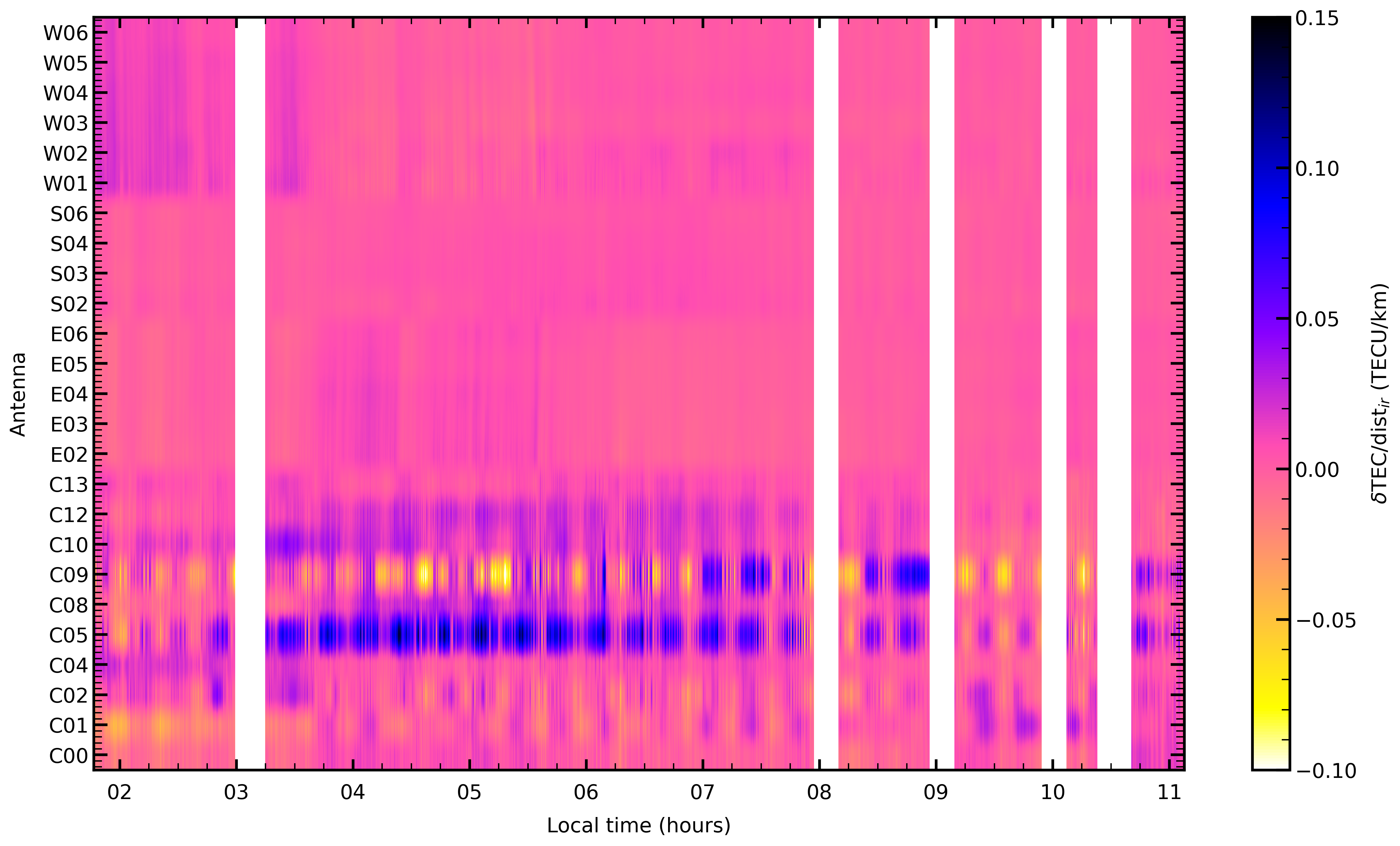}
    \caption{Geometric corrected differential TEC normalised using the projected distance (see Fig.\ref{fig:zion} middle) for a respective antenna (i) with reference antenna (r) as a function of local time. Vertical TEC gradient is nearly constant along three arms as a function of time along with their respective directions. Antenna corresponds to the central square shows the different value of gradient because of the randomness of antenna positions.}
    \label{fig:hist_tec}
\end{figure*}
\subsection{Total Electron Content (TEC)}
Following the correction described above and using equation (\ref{eq:phi2tec}), we convert the continuum-subtracted phases for each antenna and band to differential TEC, or $\delta\rm{TEC}$. Additionally, we compute the median $\delta\rm{TEC}$ using the two values (i.e., two bands) and estimate the uncertainty by computing the median absolute deviation (MAD) at each time step and the given antenna. Two nearest time steps are also included in computing MAD for each time step. By doing so, we get a total of 6 points per time step to achieve higher accuracy. In both the calculations, we used the median to suppress any spurious data points left behind. \par
The resulting $\delta\rm{TEC}$ values are plotted for each antenna in the central square in Fig.\ \ref{fig:central_sq}, the northwestern arm in Fig.\ \ref{fig:northwestern}, the northeastern arm in Fig.\ \ref{fig:northeastern}, and the southern arm in Fig.\ \ref{fig:southern} along with the MAD values (plotted in red) to demonstrate the relative accuracy to which $\delta\rm{TEC}$ is estimated. The uncertainty in $\delta\rm{TEC}$ expressed by the MAD calculations is of the order of 1\,$\times$\,$10^{-3}$ TECU, showing the remarkable ability of the GMRT to detect small fluctuations while measuring TEC. These results suggest that the variation in $\delta\rm{TEC}$ for the southern arm is more than the other two arms and the central square configuration. \par
Differential TEC along three arms (Fig.\ \ref{fig:northwestern}-\ref{fig:southern}) show a different pattern, which may be because a dominant wave is propagating with smaller-scale wave(s) or several waves are propagating in different directions. Spectral analysis of such patterns will be demonstrated in our subsequent work. Further, the larger amplitude and longer period fluctuation (known behaviour of TIDs \citealt[][]{her06}) are present throughout the observation in the southern arm, but the same characteristic fluctuations are present only during the night-time (before dawn) in the remaining two arms. Also, antennas located at different distances along the same direction or the same azimuth describe that $\delta\rm{TEC}$ appears to be proportional to the baseline length. Measuring the gradient is therefore necessary to understand the behaviour of any ionospheric phenomena.
\section{Estimate the TEC gradients}
\label{sec:tec_grad_general}
The GMRT is sensitive towards variation in TEC gradient because it only measures $\delta\rm{TEC}$ between antenna pairs, as already discussed in section \ref{sec:method}. Measuring the TEC gradient is critical for any analysis to understand TEC fluctuations, but considering the configuration of the GMRT (see Fig.\ \ref{fig:gmrt_array}), measuring the TEC gradient over the array requires some modification to the time series. Without such modifications, the set of $\delta\rm{TEC}$ time series can only be analysed spectrally like a single plane wave \citep[see][]{Jac1992A&A...257..401J} or for specific assumed models. \par
This is different from the mode of operation where sampled visibility data from radio telescopes are used to make an image. TEC gradient fluctuations can be expected as a function of actual antenna positions projected onto the ionosphere as the ionosphere changes continuously. Spatial coverage can be improved by exploiting two factors. First, the fluctuations are relatively different for each antenna as fluctuations are moving. Second, the apparent position of the target is changing because of the Earth's rotation; transforming the temporal baselines into spatial ones. However, it is not as straightforward as the Earth rotation synthesis \citep[see][]{thompson_book}. Since fluctuations in TEC gradient presumably have the distribution of directions and speed, one must analyze by decomposing the $\delta\rm{TEC}$ time series into temporal spectral modes and then study the properties for each mode to extract the direction, speed and the size of the dominant pattern(s) \citep[see][]{Helm2012RaSc...47.0L02H} over the array. A detailed analysis of such spectral modes will be demonstrated in the subsequent work. \par
\begin{figure}
    \centering
    \includegraphics[scale=0.375]{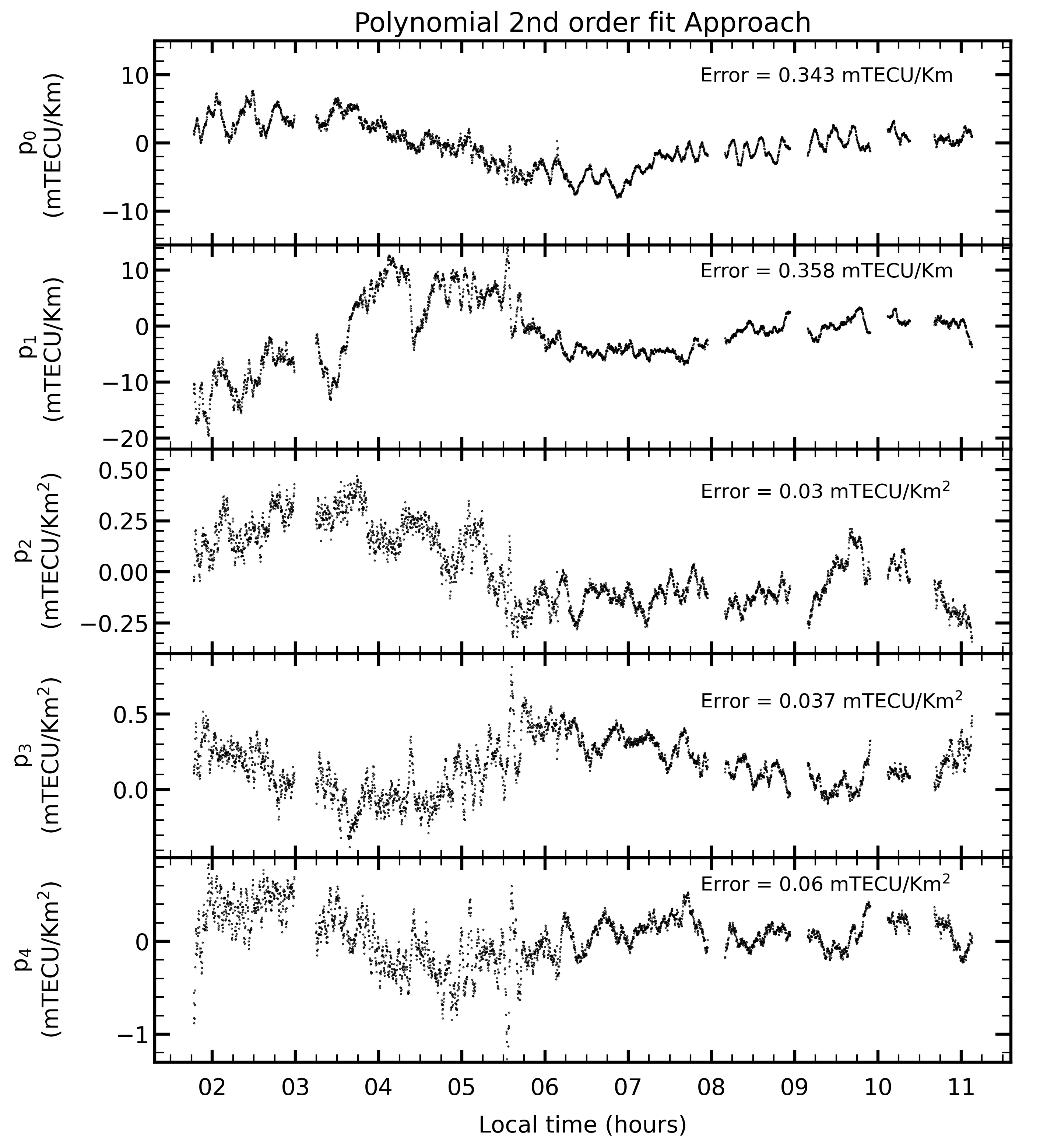}
    \caption{The fitted coefficients using the second-order polynomial equation (\ref{eq:dpoly_2nd}) as a function of local time. The coefficients were estimated using the difference between $\delta\rm{TEC}$ for each of 325 antenna pairs at each time step. The estimated standard-error for each coefficient is mentioned in the respective panel.}
    \label{fig:pfit2}
\end{figure}
Before proceeding, two geometric corrections are performed to the data such that the measured $\delta\rm{TEC}$ would correspond to the vertical $\delta\rm{TEC}$ as closely as possible. Firstly, to ensure the shape of the observed TEC surface, the array pattern of GMRT (Fig.\ \ref{fig:gmrt_array}) is projected onto the ionosphere where the antenna's line of sight passes through it (``pierce-points''). Secondly, as the apparent position of the target source 3C68.2 is changing during observation, the slant-to-vertical TEC corrections must be computed for the line of sight to the source. These corrections are relatively straightforward for parallel plane approximation. However, as 3C68.2 was very close to the horizon at the beginning and the end of the observation run, a parallel plane approximation is not applicable for all the time steps. These two geometric corrections are computed by applying a spherical model, as described in Appendix \ref{sec:appendixA}. \par
Following the spherical model, a thin shell approximation of the ionosphere surface is assumed to be located at an altitude of maximum electron density, or  ``peak height'' \citep[see][]{lan88,ma03}. Using the IRI Plas\footnote{IRI extended to Plasmasphere  \url{http://www.ionolab.org/iriplasonline/}} software, the peak height as a function of time is estimated using the latitude and longitude of the GMRT with the date and time of our observations. We re-estimate the peak heights using the pierce-points latitude and longitude. Since the results are marginally changed, we stop at one iteration. The final peak height obtained is plotted in the upper panel of Fig.\ \ref{fig:zion} (top) along with the projected distance of the farthest antennas along each arm (antennas E06, W06, and S06; see Fig.\ \ref{fig:gmrt_array}) from the array centre in Fig.\ \ref{fig:zion} (middle) and the corresponding slant-TEC to vertical-TEC corrections in Fig.\ \ref{fig:zion} (bottom). \par
The geometric corrections are applied to the $\delta\rm{TEC}$ measurements using equations \ref{eq:slant2vert} and \ref{eq:mappingfunction}. As shown in Fig. \ref{fig:hist_tec}, the vertical $\delta\mbox{TEC}$ for an antenna as a function of the local time, which is computed using the projected distance for every antenna (i) with a reference antenna (r). It can be seen that the slope is relatively constant along the three arms as a function of time along with their direction. This is because for the ionospheric wave(s) drifting over the array at constant speed, the $\delta\rm{TEC}$ will be proportional to the length of separation between the antenna when the antenna along the direction of the wave is considered. As central square antennas are randomly situated, the slope is different for each antenna along with their respective directions with the reference antenna. From Fig. \ref{fig:hist_tec}, it is evident that the strength of the temporal variations in the $\delta\rm{TEC}$ is much higher for baselines involving the central square antennas (shorter baselines) than the arm antennas (longer baseline). \par
To characterise the TEC surface over the array, we opt to measure the TEC gradient over the full array and along each arm of the GMRT using two methods as discussed below.
\subsection{Method 1: Using Polynomial Fits}
\label{sec:polynomial_fit}
\begin{figure}
    \centering
    \includegraphics[scale=0.375]{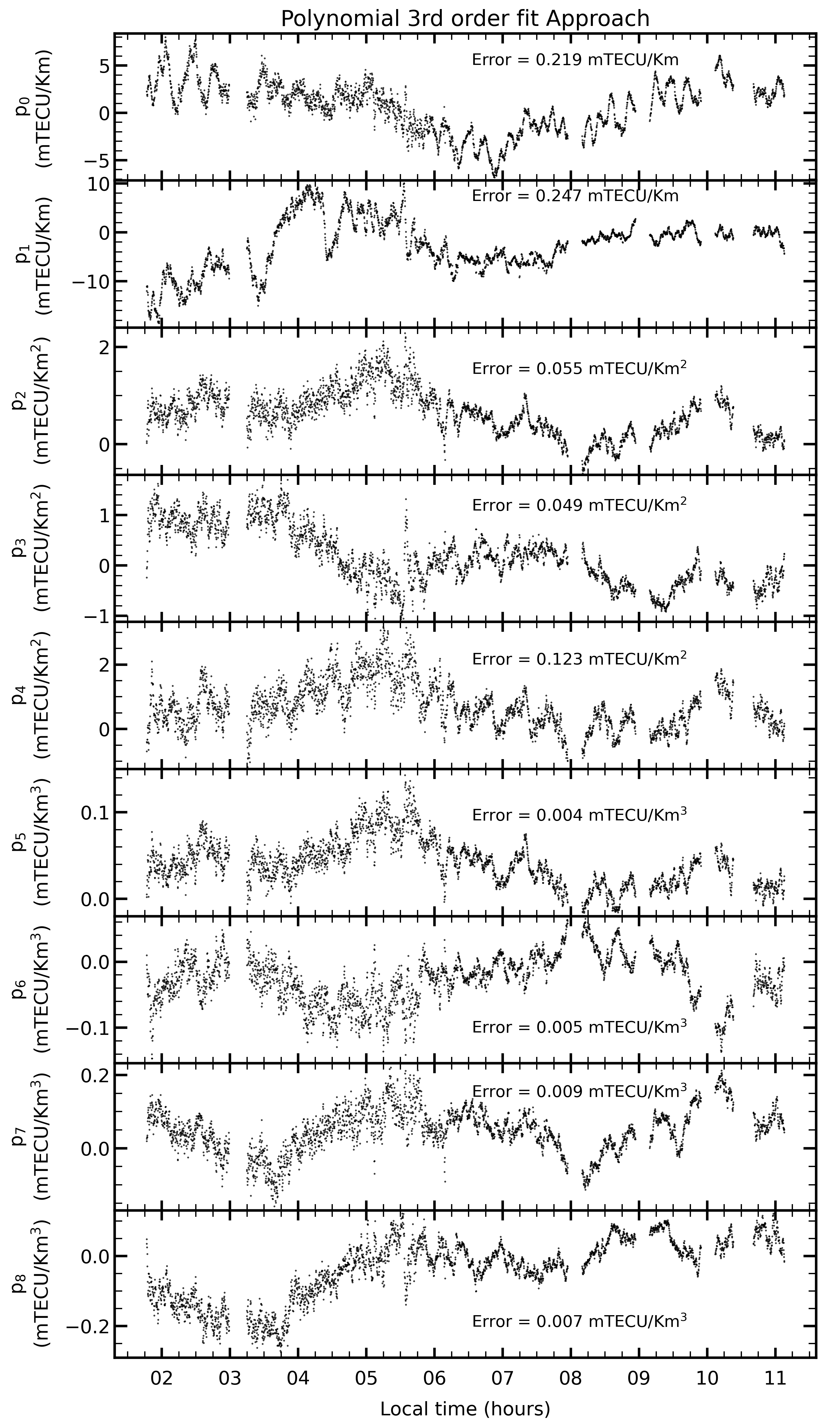}
    \caption{The fitted coefficients using the third-order polynomial equation (\ref{eq:dpoly_3rd}) as a function of local time. The coefficients were estimated using 325 antenna pairs at each time step and standard error for each coefficient is mentioned in their respective panel. Higher order coefficients are more significant during the night-time and shows the smooth variation after dawn (same as like in Fig.\ (\ref{fig:pfit2})}
    \label{fig:pfit3}
\end{figure}
To characterise the two-dimensional observed TEC gradient at each time step for any given antenna, the two geometric corrections were applied to the antenna geographical locations and the measured $\delta\rm{TEC}$ data. The two-dimensional TEC surface can be approximated using the low-order Taylor series as the size of the array is smaller than most transient ionospheric phenomena. After inspecting numerous time-steps and adequately approximate the curvature in the TEC surface, we select second and third-order terms of the two-dimensional Taylor series. These low-order Taylor series have the following form
\begin{eqnarray}
\label{eq:poly_2nd}
\mbox{TEC} \! &=& \! p_{0}\,x + p_{1}\,y + p_{2}\,x^{2} + p_{3}\,y^{2} + p_{4}\,x\,y + p_{5}
\end{eqnarray}
\begin{eqnarray}
    \label{eq:poly_3rd}
    \mbox{TEC} \! &=& \! p_{0}\,x + p_{1}\,y + p_{2}\,x^{2} + p_{3}\,y^{2} + p_{4}\,x\,y + p_{5}\,x^{3} \nonumber \\
    &+& \! p_{6}\,y^{3} + p_{7}\,x^{2}\,y + p_{8}\,x\,y^{2} + p_{9}
\end{eqnarray}
where $x$ and $y$ are the north-south and east-west antenna positions, respectively. These $x$ and $y$ are projected onto the thin-shell modelled ionosphere surface at peak height. By using the difference between $\delta\rm{TEC}$ for each of the 325 baselines at every time step, we constrain each fit's parameters to maximize the accuracy. The equations (\ref{eq:poly_2nd} and \ref{eq:poly_3rd}) that fit to the corresponding data are
\begin{eqnarray}
    \label{eq:dpoly_2nd}
    \delta \mbox{TEC}_i - \delta \mbox{TEC}_j \! &=& \! p_0\,(x_i-x_j)+p_1\,(y_i-y_j)\nonumber \\ 
    &+& \! p_2\,(x_i^2-x_j^2) + p_3\,(y_i^2-y_j^2) \nonumber \\
    &+& \! p_4\,(x_i\,y_i-x_j\,y_j)
\end{eqnarray}
\begin{eqnarray}
    \label{eq:dpoly_3rd}
    \delta \mbox{TEC}_i - \delta \mbox{TEC}_j \! &=& \! p_0\,(x_i-x_j)+p_1\,(y_i-y_j) +p_2\,(x_i^2-x_j^2) \nonumber \\
    &+& p_3\,(y_i^2-y_j^2) + p_4\,(x_i\,y_i-x_j\,y_j) \nonumber \\ 
    &+& p_5\,(x_i^3-x_j^3) + p_6\,(y_i^3-y_j^3)  \nonumber \\ 
    &+& p_7\,(x_i^2\,y_i - x_j^2\,y_j) + p_8(x_i\,y_i^2-x_j\,y_j^2)
\end{eqnarray}
where subscripts $i$ and $j$ correspond to the $i^{\mbox{\scriptsize th}}$ and $j^{\mbox{\scriptsize th}}$ antennas, respectively.
Following this, we apply standard sigma-clipping at each time step during the fitting process. This is estimated by calculating the RMS of the residuals and rejecting all baselines with absolute value greater than 3 sigma. This process is repeated until no baseline is rejected. 
Note that for a given time step, as few as zero to a maximum of 15 baselines are rejected. With this method, one can note that for every time step, the fitting is done independently, preserving the existence of any small-scale temporal/spatial TEC variations. The root-mean-square errors estimated for the second and third-order polynomial fits are 7.8\,mTECU and 5.5\,mTECU respectively. \par
\begin{figure}
    \centering
    \includegraphics[scale=0.42]{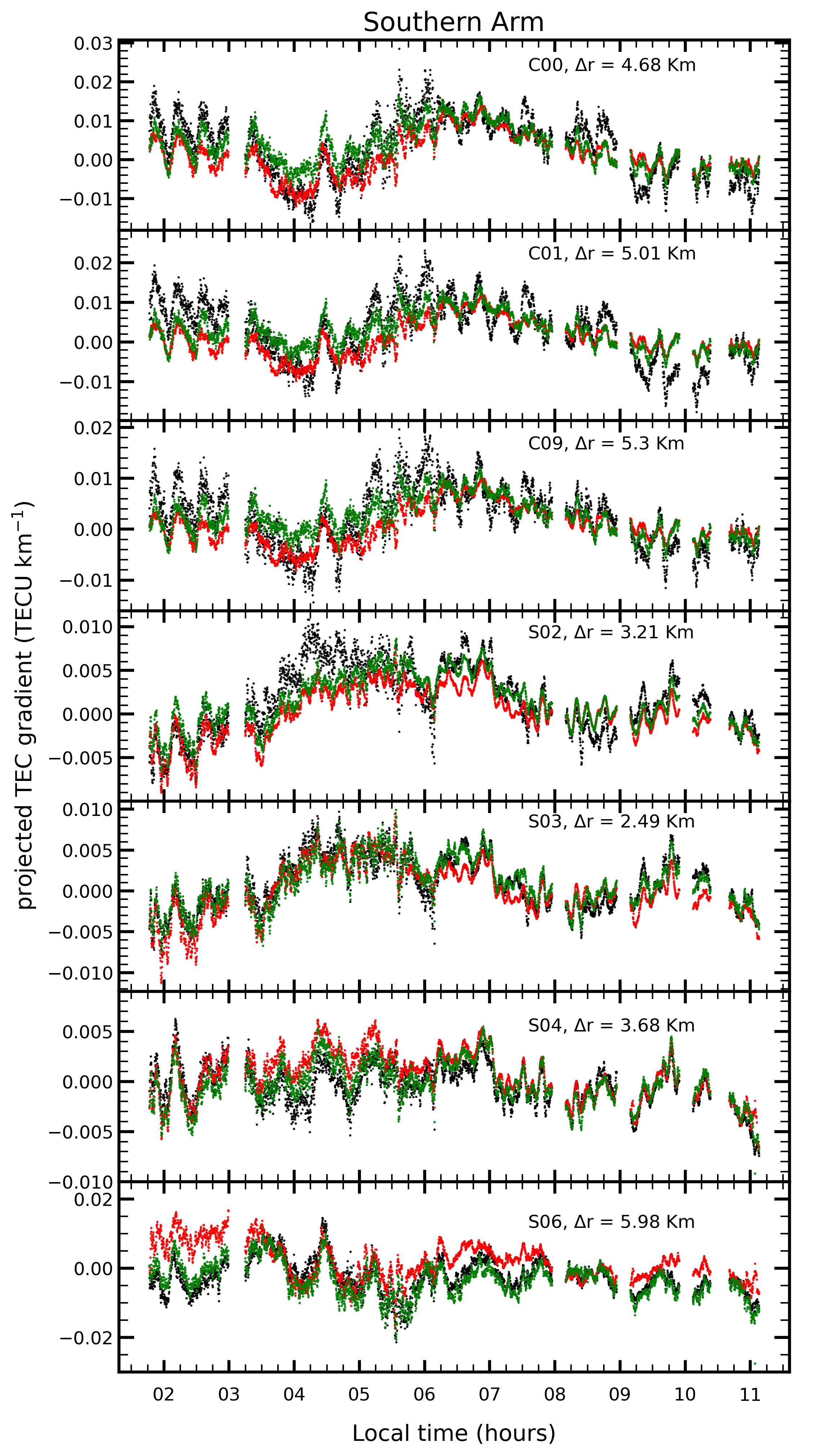}
    \caption{The projected TEC gradient time series at each antenna along the Southern Arm (black curves). Projected TEC gradient using the second-order (red) and third-order (green) polynomial fits are also plotted. In each antenna panel, the mean separation ($\Delta r$) among the antennas is printed which are used to finding the TEC gradient using the three-point Lagrangian interpolation method.}
    \label{fig:southern_grad}
\end{figure}
The polynomial coefficients are plotted as a function of local time in Fig.\ \ref{fig:pfit2} and \ref{fig:pfit3}. Additionally, the standard-error for each coefficient is mentioned in the respective panels. From the plots, one can notice that larger amplitude and longer fluctuations are visible at the beginning of the observing run, similar to the individual antenna $\delta\rm{TEC}$ time series plotted in Fig.\ \ref{fig:central_sq}--\ref{fig:southern}. However, by applying slant-TEC to the vertical-TEC correction, the fluctuations are not substantial. One can observe that the amplitude for $\rm p_1$ coefficient (a partial derivative of the TEC surface along the east-west direction) is the maximum than any other coefficient during the local night-time. Qualitatively, it is a known behaviour of medium-scale TIDs \citep[MSTIDs;][]{her06}, commonly detected around sunrise and sunset. The higher-order terms in the second-order, as well as in the third-order polynomial fit, are more significant during the night-time. The higher-order coefficients were able to capture small-scale fluctuations during this period. Fluctuations start to become prominent in the afternoon as the sun approaches the zenith. \par
\subsection{Method 2: Using Lagrangian polynomial}
\label{sec:arm_based_approach}
\begin{figure}
    \centering
    \includegraphics[scale=0.42]{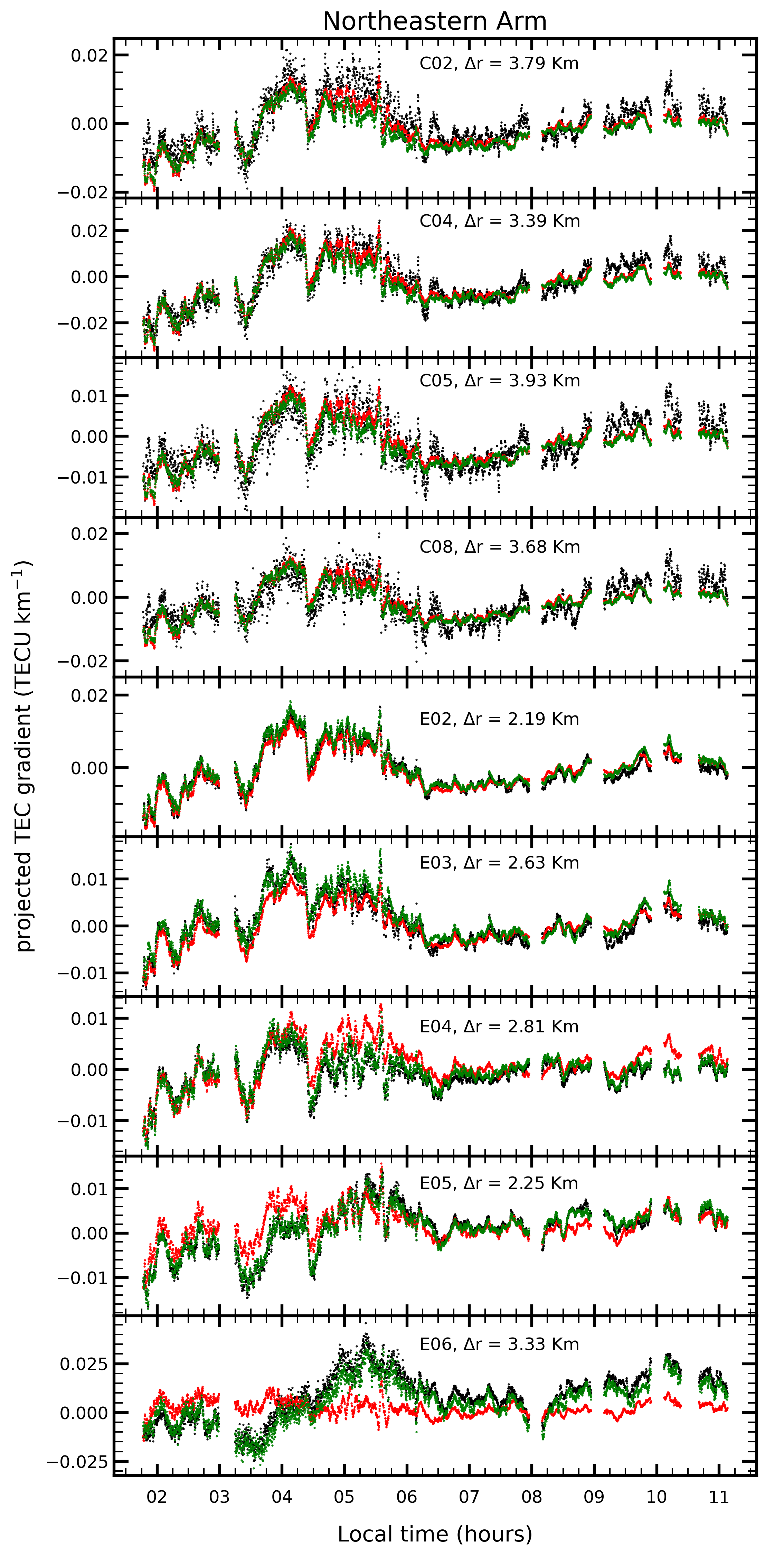}
    \caption{Same as Fig.\ \ref{fig:southern_grad} but for Northeastern Arm.}
    \label{fig:northeastern_grad}
\end{figure}
\begin{figure}
    \centering
    \includegraphics[scale=0.42]{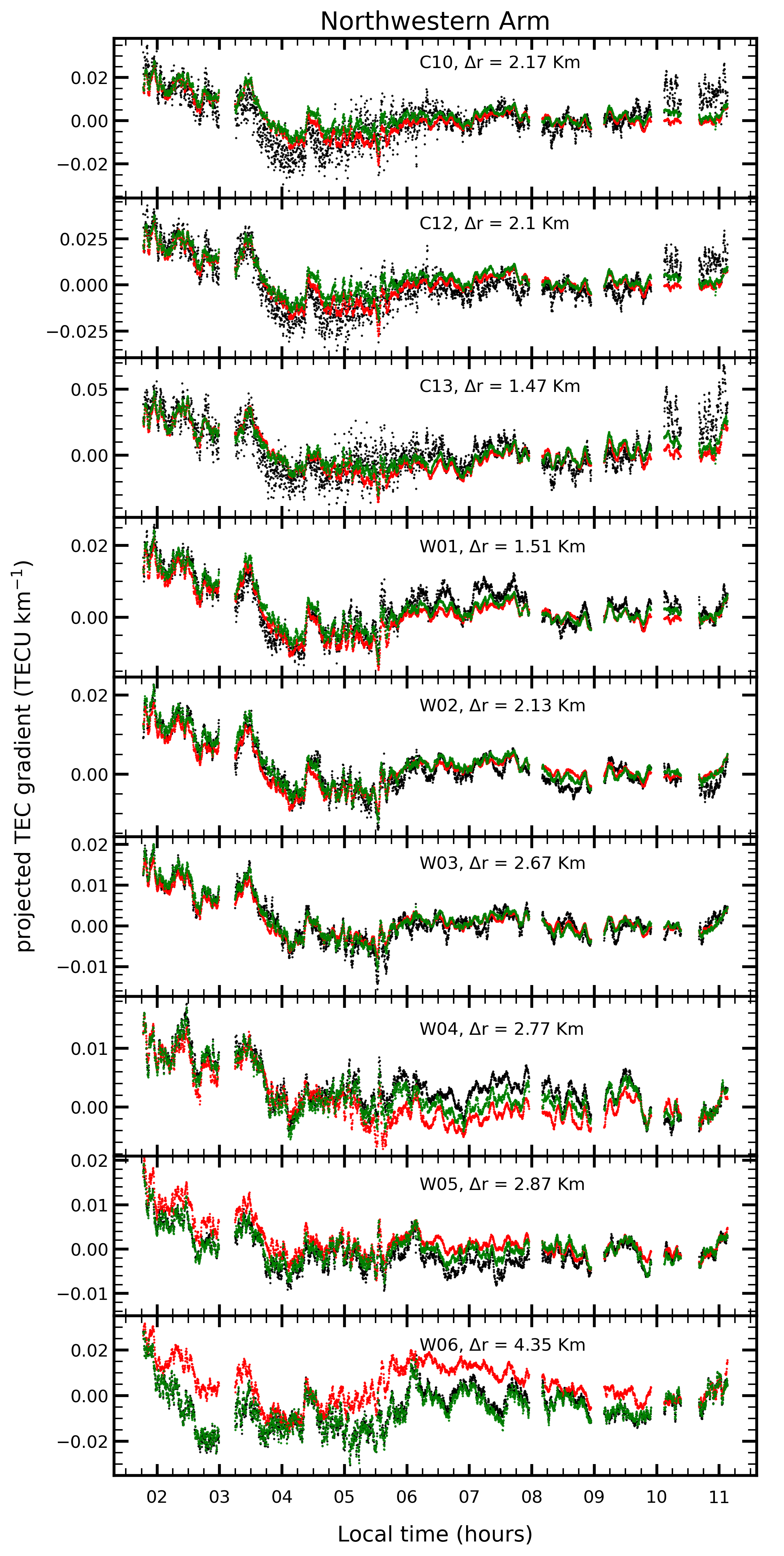}
    \caption{Same as Fig.\ \ref{fig:southern_grad} but for Northwestern Arm.}
    \label{fig:northwestern_grad}
\end{figure}
Even though the polynomial-based fitting helps us understand the two-dimensional TEC gradient variation in broader scales, this method is unable to detect small-scale fluctuations, as small as few kilometres. These measurements also show that the third-order fit is more accurate than the second-order fit in detecting these small-scale fluctuations. In principle, one could keep increasing the order of the polynomial for better results. However, it is expected that the small-scale fluctuations do not stretch over the array, which concludes that such a fitting method cannot provide an accurate description of the two-dimensional TEC gradient at any given antenna, especially the farthest ones in the arms (see Fig.\ \ref{fig:gmrt_array}). Therefore, to make full use of the data, we select an alternate method instead of going towards higher-order polynomial fits. \par
This alternate method computes the projected TEC gradient (or the spatial derivative of $\delta\rm{TEC}$) in three directions corresponding to each GMRT arm. However, in Fig.\ \ref{fig:gmrt_array}, the antennas in the central square are randomly placed and no three antennas are in a single line, thus TEC gradient cannot be computed along a single direction. To solve this, we curve-fit a straight line along with three different directions of the array and extended that line to the middle of the array. The central square antennas are considered in the respective direction of closest approach to the respective fitted lines. After doing so, the projected gradient is computed using a three-point Lagrangian interpolation for each arm's antennas at each time step. This method may be a poor predictor of the function between data points, however the accuracy at the data points will be ``perfect''. Therefore, the gradient will be more accurate at the data points (or antenna positions). \par
Considering the $\delta\rm{TEC}$ uncertainty of 1\,$\times$\,$10^{-3}$ TECU and the typical average separation between antennas of about 3 km, we get a precision of $\sim$7\,$\times$\,$10^{-4}$ TECU km$^{-1}$ for the projected TEC gradient measurements. The time series of the projected TEC gradient is depicted, following Lagrange interpolation method, for each antenna in Fig.\ \ref{fig:southern_grad}--\ref{fig:northwestern_grad} with black dots. To compare with the previously illustrated method, the projected TEC gradient computed using the second-order and third-order polynomial coefficients in Fig.\ \ref{fig:pfit2} and \ref{fig:pfit3} are also shown in red and green dots, respectively. Note that for larger amplitude and longer period of fluctuations, the third-order polynomial fit recovers much of the observed data. Furthermore, the second-order polynomial fit poorly recovers the structure for antennas beyond 6 km, from the array centre. However, during the night-time ($\leq 6$ hrs. Local time), smaller-scale fluctuations are missed by both the second-order and third-order polynomial fits, which are observed only in individual antenna TEC gradients time series computed using the Lagrangian interpolation method. These small-scale fluctuations are significant for the antennas near the centre of the array as seen in Fig.\ \ref{fig:southern_grad}--\ref{fig:northwestern_grad}.
\section{Discussion}
\label{sec:results}
Our study demonstrates the GMRT's capability to characterise different kinds of low-latitude ionospheric conditions. Long observation of 3C68.2 leads to the precision of $1\times10^{-3}$ TECU in $\delta\rm{TEC}$ measurements with $1\sigma$ uncertainty. In this study, we achieve an order of magnitude higher sensitivity in TEC measurement, compared to GPS-based \citep[see][]{MSTID_gps} measurements.
Similar studies have been previously done using telescopes situated in the mid-latitude region. \citet{Helm2012RaSc...47.0K02H} using VLA, measured $\delta\rm{TEC}$ with the precision of $\rm<\,10^{-3}$ TECU by observing a very bright source (Cygnus A). While, \citet{mev16} used LOFAR to measure $\delta\rm{TEC}$ with an accuracy of 1\,$\times$\,$10^{-3}$ TECU by utilising the data only from night-time observations of 3C196. It is important to note that both of these studies and our study is conducted in similar solar conditions and moderate amount of geomagnetic activity. 
This measurement also shows that larger amplitude and longer period fluctuations are visible within the $\delta\rm{TEC}$ data during the night-time and throughout the observation. The nature and strength of the ionospheric variability in low latitude and low solar activity, matches with other observations using the GNSS \citep[see][]{Sumanjit2020AdSpR..65..198C,Sumanjit2020AdSpR..66..895C,Deepthi2020AdSpR..65.1544A}. The polynomial-based approaches, detailed in section (\ref{sec:polynomial_fit}) was able to characterise the two-dimensional TEC gradients over the array. While insensitive to small-scale variation, this approach effectively recovers the properties associated with larger amplitude and longer period of fluctuations in a two-dimensional TEC gradient surface. The third-order polynomial fits recovers finer structures than the second-order approach during the night-time. Spectral analysis of these data can help determine the size, speed and directions of travelling ionospheric waves \citep[see][]{Helm2012RaSc...47.0L02H,Helmb2019JA..027483} and will be demonstrated in more detail in subsequent works. \par
The polynomial-based method is used previously by \citet{Helm2012RaSc...47.0K02H} to study the ionosphere structure with the VLA. The second-order polynomial fit approximated the curvature needed to recover much of the observed structure, but small-scale disturbances were missed during the night-time. Our study shows that, a second-order polynomial fit fails to recover structures for antennas which are beyond 6\,km from the array centre. However, a third-order polynomial fit recovers much of the observed structure, although small-scale fluctuations are missed during the night-time. Therefore, an additional higher-order fit is needed to approximate the curvature in the two-dimensional TEC surface to study the ionosphere over the GMRT, as it is located near the magnetic equator. \par
The second method detailed in section (\ref{sec:arm_based_approach}) numerically computes the projected TEC gradient along each arm at each antenna and time step. This method has clearly shown that small-scale disturbances are present after midnight local time that were missed within the polynomial-based methods. Following this, the GMRT can also be used to study small-scale ionospheric phenomena just like the VLA \citep[see][]{Helm2012RaSc...47.0K02H}. \par
Additionally, we demonstrate that the ionosphere over the GMRT has more structures (small-scale to large-scale) present, compared to VLA ($\sim34^{\circ}$N), LOFAR ($\sim53^{\circ}$N) and MWA ($\sim26^{\circ}$S), even during low solar activity and moderate amount of geomagnetic activity. Prior work by \citet{ADas2008RaSc...43.5002D} with the GMRT also observed the presence of large-scale periodic structures in phase during post-sunset to pre-midnight hours. While other telescope facilities can effectively observe the ionosphere in the mid-latitude region, the GMRT ($\sim19^{\circ}$N) is the only radio interferometer located between the northern crest of the EIA region and the magnetic equator, where the concentration of electron density is found to be highest, and it is roughly around 70\% of the global electron density distribution \citep[see][]{Appl1946Natur.157..691A}.\par
This unique positional advantage of the GMRT can be used to understand and characterise the ionosphere to help in low frequency radio astronomical observations \citep[see][]{Intema2009A&A...501.1185I,Intema2017A&A...598A..78I, VanW2016ApJS..223....2V}. Moreover, accurate knowledge about the ionosphere derived from radio interferometers can also be used for ionospheric studies and space weather. In the near future, such studies can also be carried out with upcoming radio telescopes like the SKA in Western Australia and South Africa.
\section*{Acknowledgements}
We thank the staff of the GMRT who have made these observations possible. GMRT is run by the National Centre for Radio Astrophysics of the Tata Institute of Fundamental Research. SM would like to thank the financial assistance from UGC. SM would further like to thank Althaf A., Ramij Raja, Arnab Chakraborty, for helpful discussions. AD would like to thank CSIR for the grant No. 03(1461)/19/EMR-II. \par
Acknowledgement is also given for the model development of the IRI-Plas model  which is freely available at \href{http://www.ionolab.org}{http://www.ionolab.org}. \par
This research made use of \texttt{ASTROPY}, a community-developed core Python package for Astronomy \citep{astropy:2013, astropy:2018}, \texttt{NUMPY} \citep{Numpy2020array}, \texttt{SCIPY} \citep{SciPy-NMeth2020}. This research also made use of \texttt{MATPLOTLIB} \citep{matplotlib07} open-source plotting packages for \texttt{PYTHON}.
%
\section*{Data Availability}
All the radio data used in this study are available in the GMRT Online Archive \href{https://naps.ncra.tifr.res.in/goa/data/search}{(https://naps.ncra.tifr.res.in/goa/data/search)} with proposal code 22\_064. 
%


\bibliographystyle{mnras}
\bibliography{main} 



\appendix

\section{Geometric Corrections}
\label{sec:appendixA}
To represent the actual condition of the ionosphere as closely as possible, two straightforward geometric corrections are applied to the antenna positions and the measured $\rm delta TEC$ values for each antenna at each time step. As the target's apparent position continuously changes throughout the observation, one cannot use a parallel-plane approximation when the source is close to the horizon during the beginning and at the end of the observation run; thus, a thin shell approximation is considered for the ionosphere at peak height. This peak height (altitude of maximum electron density, $z_{\rm ion}$) was computed by the IRI-Plas software for the time and day of the observations (see section \ref{sec:tec_grad_general} and Figure \ref{fig:zion}). The correction used is explained below. \par
First, as antenna elevation to the observed target changes throughout the observation, the antenna's geographical location must be projected onto the considered thin shell ionosphere at peak height. For a spherical shell, the line of sight for each antenna to the source that passes through the ionosphere can be defined as ``ionospheric pierce-point (IPP)'' for that antenna. Elevation and azimuth for each antenna can be easily computed using the latitude-longitude for that antenna on the ground, date and time of the observation and RA-DEC of the observed target. At a particular time, azimuth ($A_{\rm z}$) and elevation (E) of the line of sight vector from the antenna to target along with antenna latitude-longitude ($\phi_{\rm a},\lambda_{\rm a}$) is essential to calculate the IPP ($\phi_{\rm pp},\lambda_{\rm pp}$) location and are given as 
\begin{eqnarray}
    \label{eq:ipp_latlong1}
    \psi_{\rm pp}= \frac{\uppi}{2}-E-{\rm sin}^{-1}{\Big[\frac{{R_{\rm e}}  {\rm cos}{(E)}}{R_{\rm e}{+z_{\rm ion} h_{\rm I} }}\Big]}
\end{eqnarray}
\begin{eqnarray}
    \label{eq:ipp_latlong2}
    \phi_{\rm pp}=\ {\rm sin}^{-1}\ \big[{\rm sin}(\phi_{\rm a}){\rm cos}(\psi_{\rm pp}) + {\rm cos}(\phi_{\rm a}){\rm sin}(\psi_{\rm pp}) {\rm cos}(A_{\rm z})\big]
\end{eqnarray}
\begin{eqnarray}
    \label{eq:ipp_latlong3}
    \lambda_{\rm pp} = \lambda_{\rm a}+{\rm sin}^{-1}\bigg[\frac {{\rm sin}(\psi_{\rm pp})\ast {\rm sin}(A_{\rm z})}{{\rm cos}(\phi_{\rm pp})}\bigg]
\end{eqnarray}
where $R_{\rm e}$ is radius of the Earth (6371\,km). Using equations (\ref{eq:ipp_latlong1}-\ref{eq:ipp_latlong3}), the IPP latitude-longitude values from the antenna latitude-longitude are calculated. \par
Secondly, since the astronomical sources are very far away compared to the antenna distance or Earth's radius; one can assume the line of sight for two antennas from the centre of the earth to the source is approximately parallel. As the apparent position of the source is changing, so does the path length through the ionosphere. Therefore, to convert the slant-$\delta TEC$ to vertical-$\delta TEC$, we apply the mapping function $(M(E))$ \citep[see][]{jak_slant}.
\begin{eqnarray}
    \label{eq:slant2vert}
    \delta TEC_{\rm vertical} = \frac{\delta TEC_{\rm slant}}{M(E)}
\end{eqnarray}
where, 
\begin{eqnarray}
    \label{eq:mappingfunction}
    M(E)  = \Bigg[ 1 - \Big[\frac{{R_{\rm e}} . {\rm cos}{(E)}}{R_{\rm e}+z_{\rm ion}}\Big]^{2}\Bigg]^{-1/2}
\end{eqnarray}


\bsp	
\label{lastpage}
\end{document}